\documentclass[fleqn,usenatbib]{mnras}

\usepackage{newtxtext,newtxmath}
\usepackage{caption}

\usepackage[T1]{fontenc}

\DeclareRobustCommand{\VAN}[3]{#2}
\let\VANthebibliography\thebibliography
\def\thebibliography{\DeclareRobustCommand{\VAN}[3]{##3}\VANthebibliography}

\usepackage{graphicx}	
\usepackage{amsmath}	
\usepackage[dvipsnames,table]{xcolor}
\usepackage{hyperref}
\usepackage{ulem}

\title[\texttt{CatBoost} classification of \textit{Fermi}-LAT unIDs]{Classification of \textit{Fermi}-LAT unidentified gamma-ray sources using \texttt{CatBoost} gradient boosting decision trees}

\author[J. Coronado-Blázquez]{
Javier Coronado-Blázquez$^{1}$\thanks{E-mail: j.coronado.blazquez@gmail.com}
\\
$^{1}$Telefónica Tech IoT \& Big Data, Ronda de la Comunicación s/n, Madrid, Spain}

\date{Accepted 2022 July 7. Received 2022 July 5; in original form 2022 May 30}

\pubyear{2022}

\begin{document}
\label{firstpage}
\pagerange{\pageref{firstpage}--\pageref{lastpage}}
\maketitle

\begin{abstract}
The latest \textit{Fermi}-LAT gamma-ray catalog, 4FGL--DR3, presents a large fraction of sources without clear association to known counterparts, i.e., unidentified sources (unIDs). In this paper, we aim to classify them using machine learning algorithms, which are trained with the spectral characteristics of associated sources to predict the class of the unID population. With the state-of-the-art \texttt{CatBoost} algorithm, based on gradient boosting decision trees, we are able to reach a 67\% accuracy on a 23--class dataset. Removing a single of these classes --blazars of uncertain type-- increases the accuracy to 81\%. If interested only in a binary AGN/pulsar distinction, the model accuracy is boosted up to 99\%. Additionally, we perform an unsupervised search among both known and unID population, and try to predict the number of clusters of similar sources, without prior knowledge of their classes. The full code used to perform all calculations is provided as an interactive Python notebook.
\end{abstract}




\section{Introduction}

Since its launch on June 11, 2008, the Large Area Telescope on board the NASA \textit{Fermi Gamma-ray Space Telescope} (\textit{Fermi}-LAT) has been surveying the sky searching for gamma-ray sources \cite{fermi_instrument_paper}. The \textit{Fermi}-LAT is a pair conversion telescope designed to observe the energy band from $\sim$20 MeV to more than 300 GeV. Several point-source {\it Fermi}-LAT catalogs have been released and contain hundreds to thousands of gamma-ray objects, many of them previously unknown \cite{3FGL_paper,2FHL_paper,3FHL_paper}. The difference between such catalogs lies in the different energy ranges and exposure times, and the usage of the best available astrophysical diffuse emission model and instrumental response functions (IRFs) at the time.

The gamma-ray sky can be decomposed in several pieces, such as the interstellar gamma-ray emission, the isotropic gamma-ray background (IGRB) and individual point-like and extended sources. Both the Galactic diffuse emission and the IGRB are the main difficulties to detect point-sources, due to spectral confusion and photon spill over, given the poor angular resolution of the LAT \cite{2016ApJS..223...26A}. This effect is especially strong near the Galactic plane, where the emission of the Milky Way (MW) dominates any other component\footnote{In fact, this diffuse emission is responsible for $\sim80\%$ of all photons detected by the LAT \cite{Ackermann_2012}.}.

Contrary to cosmic rays (CRs), the direction of propagation of gamma rays is (almost) unperturbed, and a precise location of sources can be made. These sources, which are not transient phenomena (although they can undergo flaring periods), are associated with astrophysical objects, which have mechanisms that accelerate particles up to these energies. These objects can be located in the MW or outside it.

Associating a gamma-ray source to a known astronomical object is not trivial, and requires a careful, multiwavelength analysis with several instruments \cite{2017ApJ...838..139S}. The LAT is able to detect a rich variety of sources, being active galactic nuclei (AGN) the most frequent.

The current picture of the AGN landscape is a unified scheme, depending on the emission in the radio band and the existence of jets, as well as their orientation towards the Earth. This classification method gives rise to objects such as blazars (BLL), quasars, or Seyfert galaxies \cite{Urry_1995}. In particular BLLs, which are AGNs with a relativistic jet pointing directly towards the Earth, are the most numerous among all known gamma-ray sources.

Concerning Galactic sources, there is a much richer variety of objects, as their intrinsic low gamma-ray luminosity can be detected from the Earth, given their distances. Among them, we may find pulsars (PSRs), pulsar wind nebul\ae~(PWNe), supernova remnants (SNRs), star-forming regions, globular clusters, high- and low-mass X-ray binaries, binary star systems, and nov\ae~\cite{Abdollahi_2020}. Among these, PSRs, PWNe, SNRs and binaries are the most common Galactic gamma-ray sources, as well as primary sources of accelerated CRs \cite{1995Natur.378..255K}. This particle acceleration produces a gamma-ray spectrum which is typically best-fit by a power law --reflecting the non-thermal production of such energetic photons. Nevertheless, PSRs, SNRs, and PWNe present a curved spectrum \cite{2001AIPC..558..115H, Takata_2015}.

PSRs also present a very quickly-rotating subclass, with periods around the millisecond, and accordingly known as MilliSecond Pulsars (MSPs). These form when an old PSR is in a binary system, with a much less massive companion star, starts accreting material from it until it reignites with a quicker pulsation period. This leads to luminosities up to 20 times larger than in a young PSR, due to their fast rotation \cite{Calore_2014}. Due to their age and reignition kick velocity, MSPs tend to appear at higher Galactic latitudes than younger counterparts \cite{Cordes_1997}.

All the source information is ultimately condensed in gamma-ray source catalogs\footnote{Every LAT catalog ever published can be found \href{https://fermi.gsfc.nasa.gov/ssc/data/access/lat/}{here}.}. Depending on the instrument, the nomenclature may vary, but it is typically an alphanumeric chain which marks the instrument and data release, followed by the equatorial coordinates in format J2000, i.e., \texttt{JHHMM.m+DDMMa}, although notable sources may be named differently. This allows a quick cross-correlation between multiwavelength catalogs. Additional data such as detection significance, flux, positional error, spectral best-fit model parameters, variability and association(s) is provided.

All the gamma-ray source catalogs share a common feature: the large fraction of unidentified sources (unIDs). Indeed, ca. 1/3 of objects are of unknown nature. This is also the typical value in the TeV regime, where 26\% of all sources are currently unassociated \footnote{See \href{http://tevcat2.uchicago.edu/}{TeVCat} online tool.}. The knowledge of the source class is critical to understand the underlying gamma-ray emission physics, as well as perform population studies \cite{2019JCAP...07..020C, Orlando:2021S8}.

In recent years, machine learning has become a widely used tool in data science for tasks such as regression, clustering, dimensionality reduction, feature extraction, and classification. Indeed, several well-established algorithms such as Logistic Regression (LR), K-Nearest Neighbours (KNN), Naive Bayes (NB), Support Vector Machines (SVMs), Artificial Neural Networks (ANN), Decision Trees (DT) or Random Forests (RF) are widely used for the latter (see, e.g., \cite{8862451, Sarker_2021} for a review on these techniques). More recent implementations include eXtreme Gradient Boost (\texttt{XGBoost}) \cite{2016arXiv160302754C}, \texttt{LightGBM} \cite{Ke2017LightGBMAH}, and \texttt{CatBoost}\footnote{\url{https://catboost.ai/}} \cite{2017arXiv170609516P}, currently the state-of-the-art in classification.

Some of the ``classic'' algorithms have been used for a classification of unIDs within gamma-ray catalogs (e.g., ANN by \cite{3fglzoo_paper}, RF and LR by \cite{2016ApJ...820....8S}), but only to distinguish between two different classes, such as PSR vs. AGN, or BLL vs. FSRQ.

In this paper, we perform a machine learning classification of the unIDs with three models: i) considering all 23 different classes; ii) excluding a single class, which induces confusion; and iii) establishing a binary PSR-/AGN-like classification. Additionally, we perform an unsupervised clustering on both known and unID populations. This will allow to better understand the underlying population, as well as compare the statistical similarity between associated and unID sources.

The paper is structured as follows: In Section \ref{sec:4fgl-dr3}, we describe the latest LAT catalog, and select the features to take into account. In Section \ref{sec:classification}, we show the results of the classification with \texttt{CatBoost} within known sources as validation, and apply it to the unID population. In Section \ref{sec:clustering}, we perform the unsupervised clustering among the known and unID population. We conclude in Section \ref{sec:conclusion}.

\section{The 4FGL--DR3 catalog}
\label{sec:4fgl-dr3}
The 4FGL catalog was released in 2019, covering 8 years of LAT data (2008-2018) \cite{Abdollahi_2020}. A second release, 4FGL--DR2, was published a year later with two additional years of data \cite{2020arXiv200511208B}, while a third one --and latest at this writing--, 4FGL--DR3, was published in January, 2022 covering 12 years of LAT operations \cite{2022arXiv220111184F}. It is therefore the most complete compendium of gamma-ray sources, as the LAT, being a space-borne telescope, covers the whole sky. We devote the interested reader to the \href{https://fermi.gsfc.nasa.gov/ssc/data/access/lat/12yr_catalog/}{LAT data repository} for technical details on the construction of the catalog.

The 4FGL--DR3 contains 6659 individual sources, from which 2296 (34\%) are unIDs. We obtain the table as an Excel-compatible file from \href{https://heasarc.gsfc.nasa.gov/db-perl/W3Browse/w3table.pl?tablehead=name\%3Dfermilpsc&Action=More+Options}{here}, where we exclude columns such as RA, DEC (yet, keeping GLON, GLAT), the alternative and extended gamma-ray names, as well as additional information of association (except \texttt{'source\_type'}, which is out target variable).

A total of 50 columns are present in our data, with information such as the flux with uncertainty, detection significance, best-fit spectrum type, best-fit parameters, variability and other spectral parameters. Some columns, devoted to peak time range and energies for flaring sources, are null in most sources, and therefore are removed from the sample. These are the \texttt{['significance\_peak', 'flux\_peak','flux\_peak\_error', 'time\_peak', 'time\_peak\_interval']} columns.

Given the distribution of null values, we decide to also remove the following columns: \texttt{['lp\_beta\_error', 'lp\_epeak', 'lp\_epeak\_error', 'plec\_exp\_index\_error', 'plec\_epeak', 'plec\_epeak\_error']}. In addition, we remove the \texttt{'name'} column and the \texttt{['semi\_major\_axis\_68', 'semi\_minor\_axis\_68', 'position\_angle\_68']} columns, as they are redundant with the 68\% values, also present in the catalog.

Therefore, we use 39 features as input for the classifier. Regarding the source class, encoded in the \texttt{source\_type} column, the catalogs distinguishes between upper- and lower-case associations depending on them being firm associations or identifications. For our purposes, we homogenize the nomenclature to lower-case classes (e.g., BLL and bll classes are merged as bll). Also, sources labeled as unknown ('unk') and without source class ('NaN') are merged in 'unk'.

\begin{figure}
	\includegraphics[width=\linewidth]{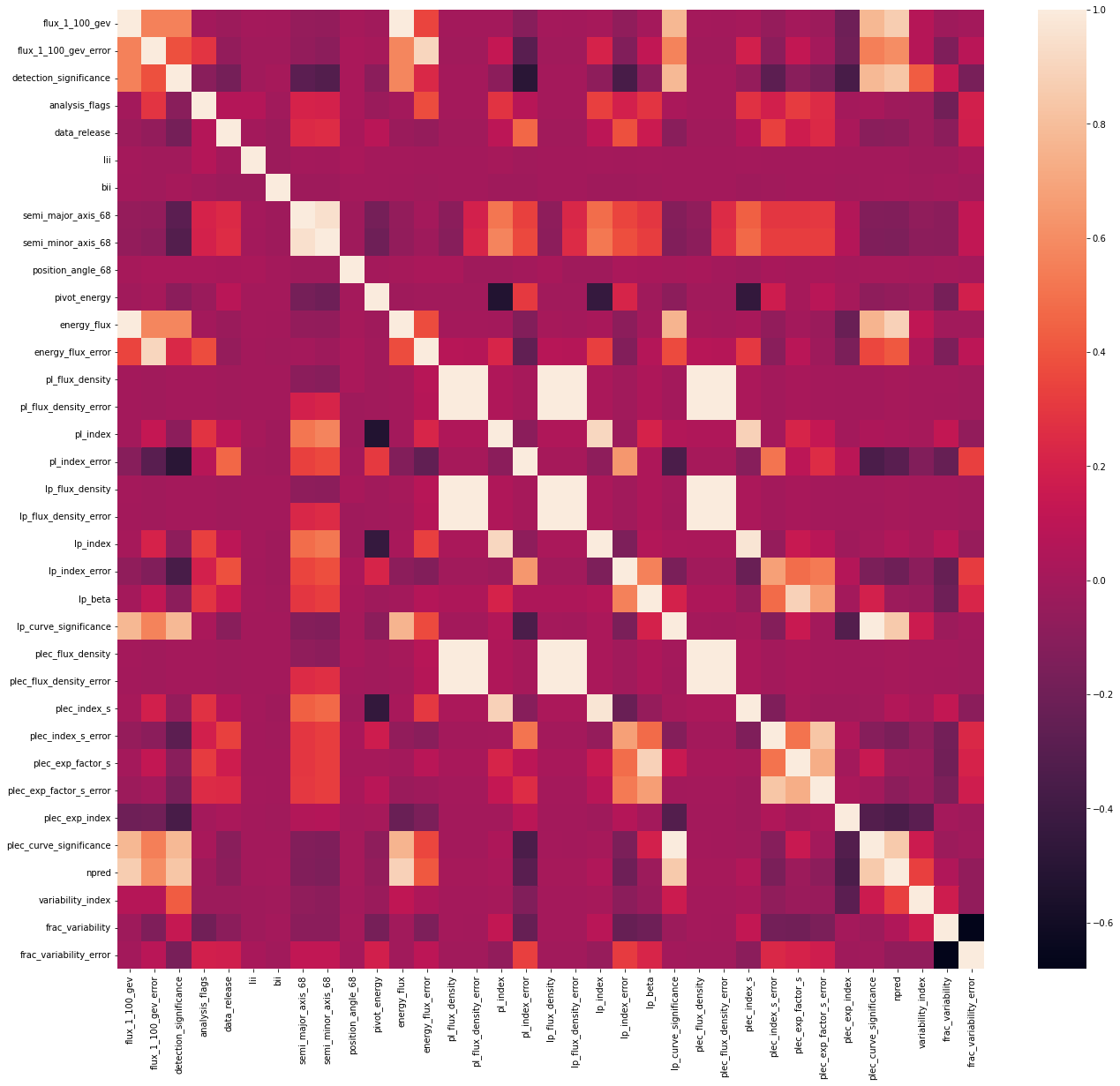}
    \caption{Pearson correlation matrix of the selected numerical variables of the 4FGL--DR3 input data. Brighter (darker) colors indicate positive (negative) correlations. A higher resolution version of all images in this paper can be generated with the provided notebook.}
    \label{fig:correlation}
\end{figure}

We also perform a study of correlations between numerical variables via Pearson correlation, as shown in Figure \ref{fig:correlation}. No significant correlations are found besides the spectral parameters and derived quantities.

\section{Classification of gamma-ray sources with \texttt{CatBoost} algorithm}
\label{sec:classification}

To shed some light on the nature of 4FGL-DR3 unIDs, we rely on \texttt{CatBoost} algorithm for classification \cite{2017arXiv170609516P}. \texttt{CatBoost} is based on a boosted decision tree with hyperparameter auto-tuning, native support for GPU executions and categorical labels (e.g., \texttt{SpectrumType}). At this writing it is the state-of-the-art algorithm for multiclass classification (see \href{https://catboost.ai/#benchmark}{this page} for more details), improving the results of \texttt{XGBoost} and \texttt{LightGBM} algorithms\footnote{In addition, we tried some ``classical'' models such as LR and kernel SVM. The results of such algorithms are not as competitive as \texttt{CatBoost}.}.

\subsection{Validation on identified sources}
\label{sec:maths}
To make predictions of the source class among the unIDs, first we must estimate the performance of the classifier in the identified sources, where we know the actual label. To do so, we generate a subset of sources taking into account only the ones with \texttt{source\_type != 'unk'}, resulting in 4363 labelled sources. As mentioned, categorical labels are automatically handled by \texttt{CatBoost} and there is no need to perform a transformation to put them as dummy variables.

As input we use 23 different classes: \texttt{['bcu' 'bll' 'fsrq' 'spp' 'psr' 'rdg' 'agn' 'msp' 'glc' 'snr' 'gal' 'sbg'
 'sfr' 'bin' 'hmb' 'nlsy1' 'lmb' 'nov' 'css' 'pwn'
 'ssrq' 'sey' 'gc']}. We refer the reader to the \href{https://heasarc.gsfc.nasa.gov/W3Browse/fermi/fermilpsc.html#source_type}{Fermi webpage} for a detailed explanation of each of these classes.

A 75/25\% train/test split is performed on the subset of known sources. With the \texttt{CatBoostClassifier} class, we train the algorithm on the train set, this is, with knowledge of the class labels, and evaluate its performance in the test set, without prior knowledge on the source type. We also perform a cross-validation on 5 folds of the test set, and run the algorithm with 1000 iterations in each fold.

\begin{figure}
	\includegraphics[width=\linewidth]{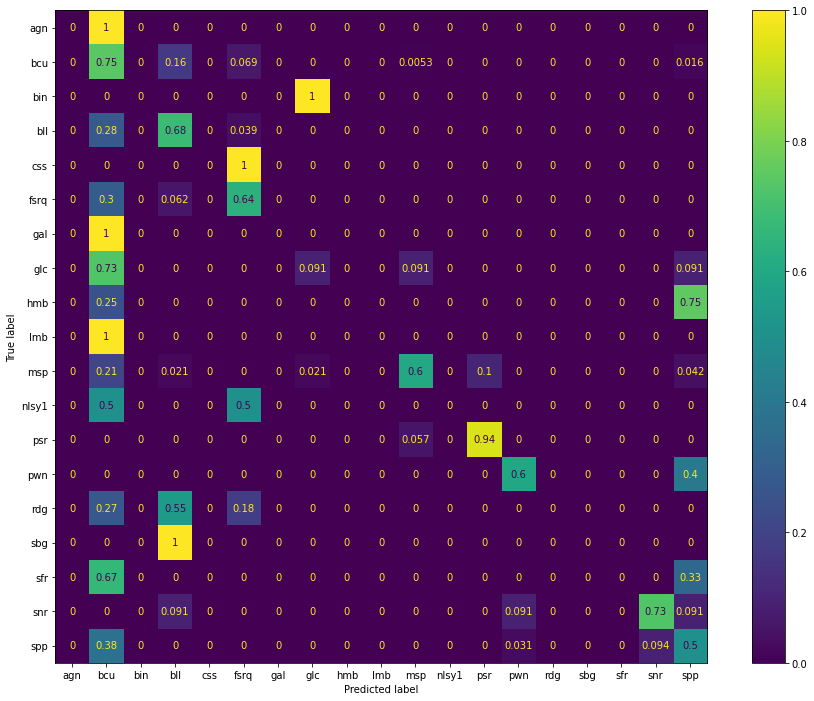}
    \caption{Normalized confusion matrix of the classifier performance on the test set of known sources, for all 23 source types. The diagonal indicates a match between the predicted and true label, while off-diagonal elements indicate a mismatch.}
    \label{fig:confusion_matrix}
\end{figure}

\texttt{CatBoost} obtains a $66.81\pm 1.18$\% accuracy on the 5--fold test set\footnote{If an algorithm randomly assigned a class, its expected average accuracy would be 100\%/23 classes=4.3\%.}. In Figure \ref{fig:confusion_matrix}, we show the confusion matrix of an average fold. Although the normalization may be misleading, with several "1s" (100\%) appearing off-diagonal, this is due to the fact that many of these "1s" reflect scarce source classes with just one or two members in the test set. If we plotted the confusion matrix with absolute numbers, it would be dominated by the bll class, as the catalog is, making very difficult to distinguish the non-zero off-diagonal elements. Nevertheless, we refer the reader to the provided notebook, where the option \texttt{normalize='true'} can be switched off in the confusion matrices generation.

We highlight the fact that most of the misclassified sources are labeled as BCUs, which are in fact a misclassified source class itself, as they are blazars of unknown nature \cite{Abdollahi_2020}. The performance of the algorithm will improve by taking into account this fact, which we will do in two different ways.

\begin{figure}
	\includegraphics[width=\linewidth]{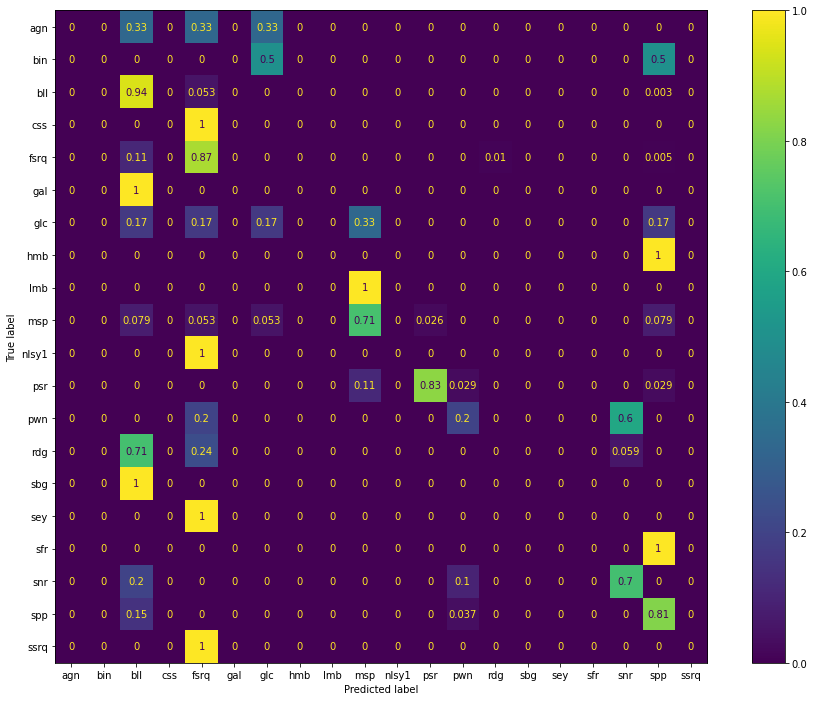}
    \caption{Normalized confusion matrix of the classifier performance on the test set of known sources, for the 22-class dataset, ignoring BCUs for inherently being a misclassified source type and the main source of confusion in the full, 23-types classification (see Figure \ref{fig:confusion_matrix}).}
    \label{fig:confusion_matrix_no_bcu}
\end{figure}

The first strategy consists on ignoring the BCU class. This will reduce the sample of known sources from 4363 down to 2763 (37\% of rejections), while retaining 22 classes. By repeating the steps in the previous, full 23-class model, we now reach a $80.70\pm2.41$\% accuracy in the test set with 5--fold cross-validation and 1000 iterations. We plot the confusion matrix in Figure \ref{fig:confusion_matrix_no_bcu}, where the main source of confusion is now the bll class. Nevertheless, this model performs a factor $\sim18$ better than a random classifier, and can provide hints for unIDs specific classes.

As a second approach, we group the sources in two general categories, g-AGNs (containing bcu, bll, fsrq, agn classes) and g-PSR (composed of psr and msp source types)\footnote{As in the rest of the paper, we will not distinguish upper- and lower-case classes.}. The distinction makes sense, as AGNs tend to present power-law-like spectra, while pulsars feature highly curved spectra with a sharp cutoff at a few GeV \cite{Takata_2015}.

This is a very similar approach to that of \cite{2016ApJ...820....8S}, which aimed to search for possible pulsars among unIDs. This way, we ignore the rest of source types, reducing the sample of classified sources from 4363 to 4041. With this compromise, by rejecting just 7\% of the associated sources, we can perform a binary classification to distinguish between pulsars and AGNs.

\begin{figure}
	\includegraphics[width=0.9\linewidth]{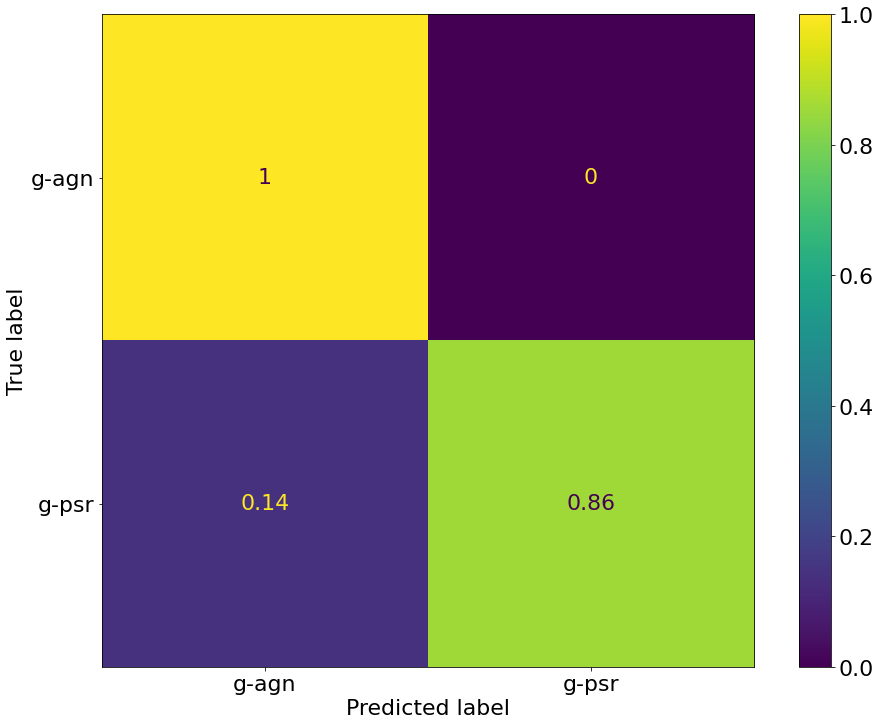}
    \caption{Normalized confusion matrix of the classifier performance on the test set of known sources, for the binary aggrupation of AGN-like (g-agn, composed of bcu, bll, fsrq, agn classes) and pulsar-like (g-psr, composed of psr and msp classes). Note the algorithm perfectly classifies the AGN-like sources.}
    \label{fig:confusion_matrix_grouped}
\end{figure}

The results of this model using \texttt{CatBoost} is an important improvement with respect to both the 23- and 22- types classification, reaching a test accuracy of $99.21\pm0.19$\% over a 5--fold cross-validation and 1000 iterations. The confusion matrix is shown in Figure \ref{fig:confusion_matrix_grouped}.

Depending on the requested level of detail of the source class, we can opt for a broad, binary classification between PSR-like and AGN-like sources, a detailed 22-class, or a 23-class (including BCUs) model, reproducing the original catalog. The assumptions and reached accuracies in the test set are summarized in Table \ref{tab:summary_models}.

\begin{table}
\centering
\begin{tabular}{ |c|c|c|  }
\hline
\rowcolor[gray]{.8} 
\multicolumn{3}{|c|}{Summary of trained models on known test sources} \\
\multicolumn{1}{|c|}{Classes} & Definition & Test set accuracy \\
\hline
\rowcolor[gray]{.9} 
23 & All classes & $66.81\pm1.18$\%\\
22 & No BCUs & $80.70\pm2.41$\%\\
\rowcolor[gray]{.9} 
2 & PSR-/AGN-like & $99.21\pm0.19$\%\\
\hline
\end{tabular}
\caption{Summary of the three models trained with \texttt{CatBoost}, according to the number of classes considered. The reported test set accuracy is computed with a 5--fold cross-validation on a 25\% test split and 1000 iterations.}
\label{tab:summary_models}
\end{table}

With these three models, we can now predict the classes of the unIDs, and study the distribution and differences between them.

\subsection{Prediction of unIDs classes}
\label{sec:prediction}
We predict the labels of each of the classes in the three models previously introduced. We use the \texttt{predict} method based on the trained data for each model. In the following, we discuss the results for each of them, according to the output distribution,

\begin{center}
\begin{tabular}{c|c}
\multicolumn{2}{c}{23-class model} \\
\hline
Class & Count\\
\hline
\hline
bcu & 1516\\
spp & 479\\
bll & 133\\
fsrq & 56\\
msp & 50\\
glc & 41\\
snr & 16\\
pwn & 3\\
rdg & 1\\
\end{tabular}
\hspace{0.1in}
\begin{tabular}{c|c}
\multicolumn{2}{c}{22-class model} \\
\hline
Class & Count\\
\hline
\hline
spp & 812\\
bll & 752\\
fsrq & 460\\
msp & 132\\
glc & 113\\
snr & 22\\
pwn & 3\\
rdg & 2\\
&\\
\end{tabular}
\hspace{0.1in}
\begin{tabular}{c|c}
\multicolumn{2}{c}{2-class model} \\
\hline
Class & Count\\
\hline
\hline
AGN-like & 2189\\
PSR-like & 107\\
& \\
& \\
& \\
& \\
& \\
&\\
&\\
\end{tabular}
\captionof{table}{Distribution of source predicted classes for each of the three considered models.}
\end{center}

The prediction of the two first models, with 23 and 22 classes, yields spp as the most or second most predicted source among unIDs, mounting up to 21 and 35\% of the whole unID class prediction, respectively. In the 23--class model, spp represents a 3\% of the total known sources, and a 4\% in the 22-class model. This class is defined in the catalog as ``special case -- potential association with SNR or PWN'' \cite{Abdollahi_2020}. Therefore, it is difficult to predict the actual class of these unIDs rather than pinning them down to these possibilities.

While we must stress that associated test set accuracies for these two models are 67 and 81\%, it is interesting to note that their confusion matrices (Figures \ref{fig:confusion_matrix} and \ref{fig:confusion_matrix_no_bcu}) do not present Type I errors for the spp class except for marginal source types, i.e., this class is not overrepresented in the output of the known sample run.

The rest of sources are associated with AGNs or pulsars, with specific sub-types. Classes which are underrepresented in the associated dataset, such as rdg, are also very scarce in the predicted set, while many are not predicted at all.

In the case of the 2-class model, it reproduces the distribution of the associated dataset, where the g-PSR/g-AGN proportion is 5/95\%, while in the case of the prediction it is 7/93\%. The result of this model can be taken more confidently than in the other models, as the average test set accuracy was >99\%. We must note that, attending to the confusion matrix in Figure \ref{fig:confusion_matrix_grouped}, the algorithm perfectly classified all AGNs but featured some g-PSR misclassified as g-AGN. Thus, we should expect the AGN-like predicted class to contain a few misclassified PSR-like objects, but none or marginal PSR predictions actually being AGNs.

\begin{figure}
	\includegraphics[width=1\linewidth]{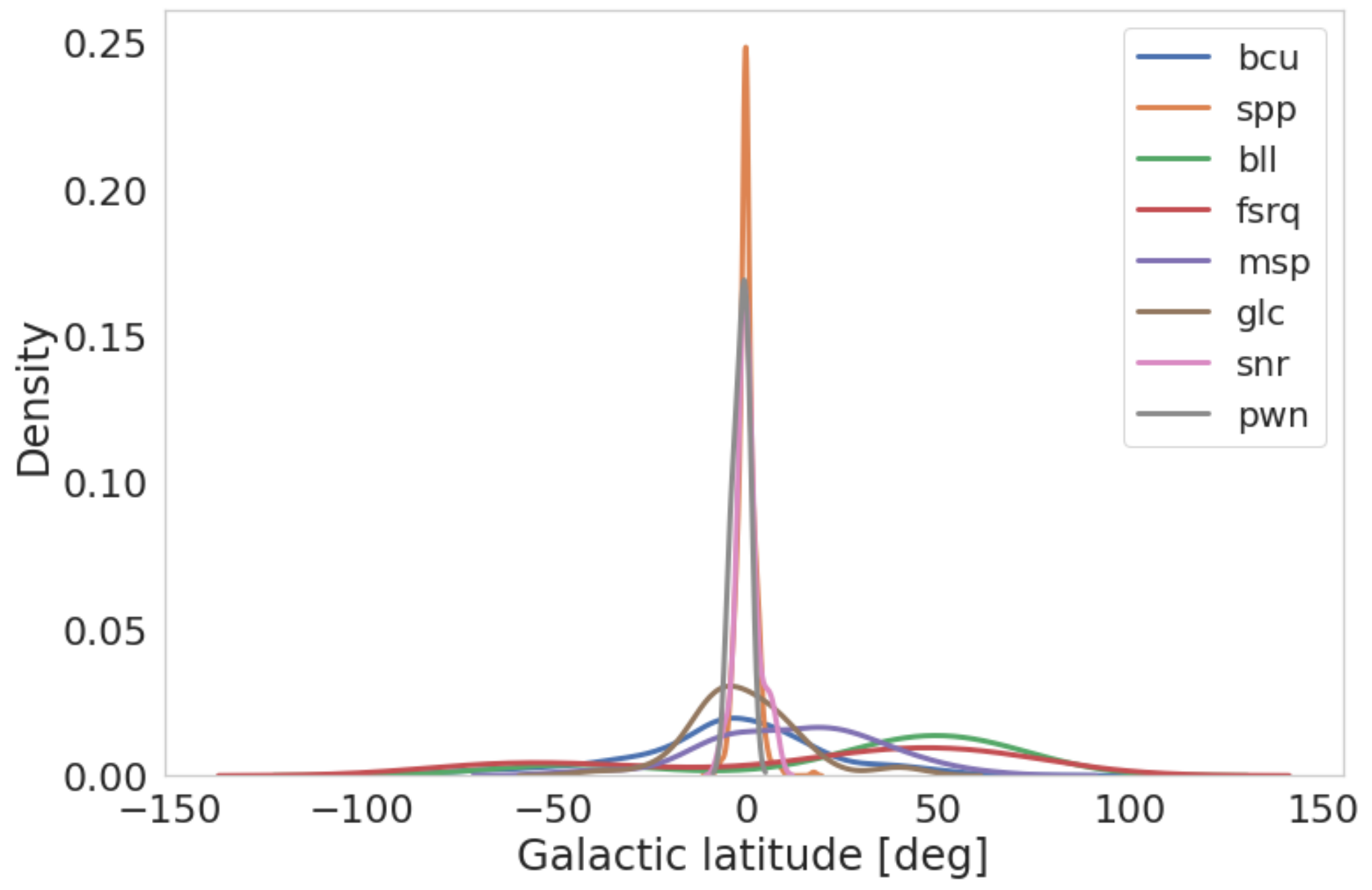}
	\vfill
	\includegraphics[width=1\linewidth]{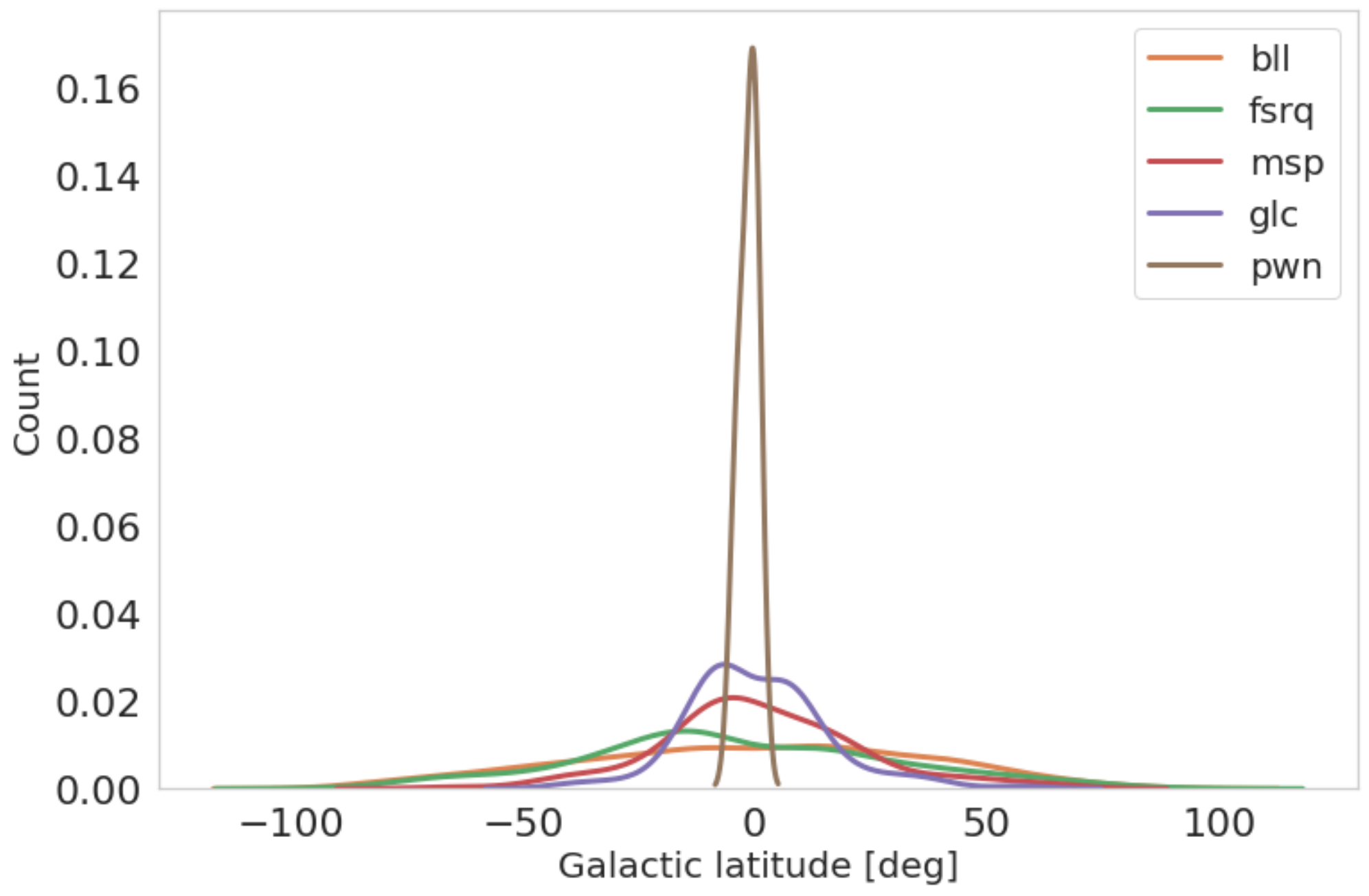}
	\vfill
	\includegraphics[width=1\linewidth]{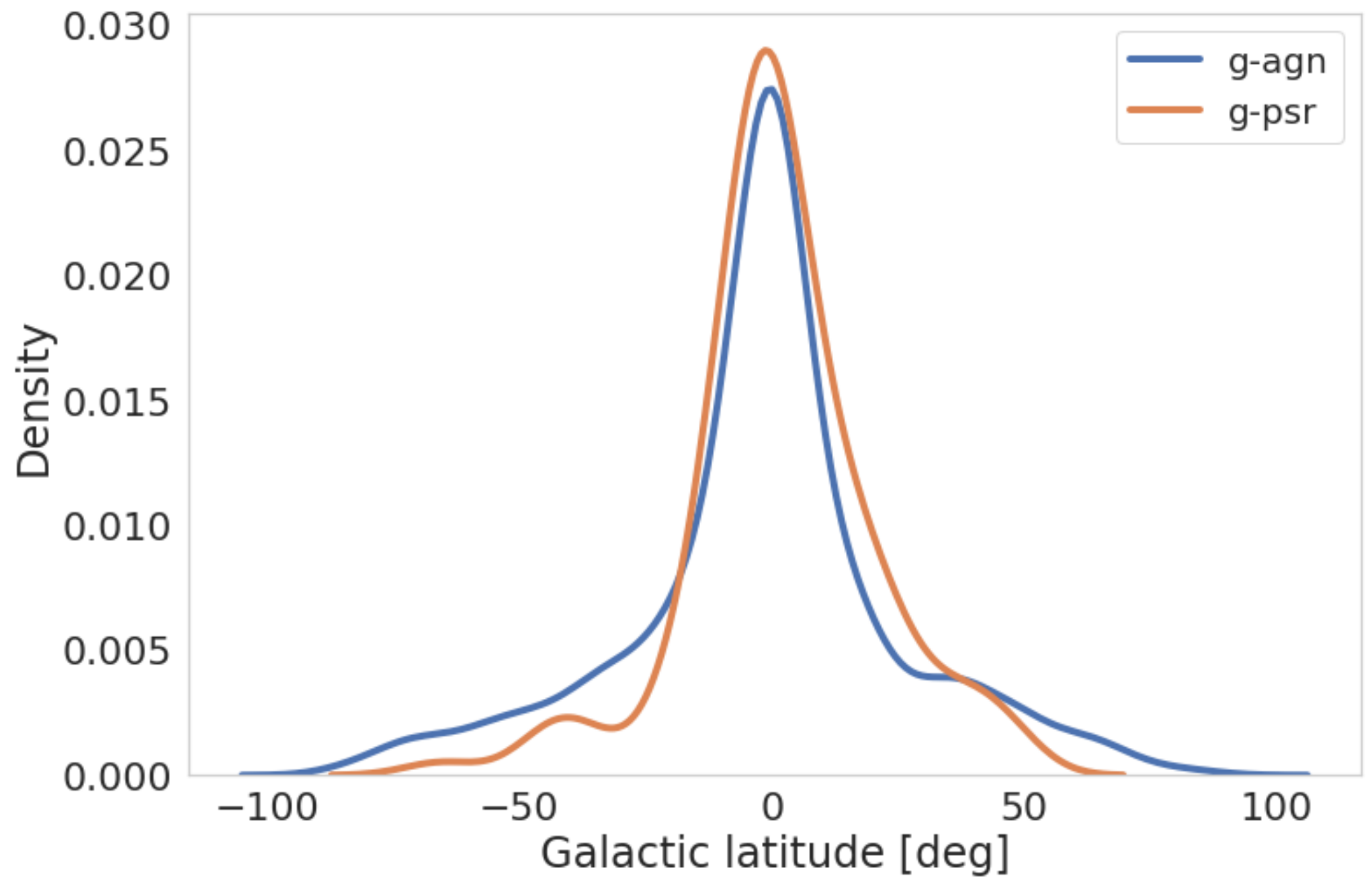}
   \caption{Galactic latitude probability distribution of the predicted classes in each of the three models. \textbf{Top row :} 23--class model. \textbf{Middle row:} 22--class model. \textbf{Bottom row:} 2--class model. As the Figure depicts the mathematical probabilistic density function, it may overcome the physical range [-90, 90].}
    \label{fig:pred_latitudes}
\end{figure}

\begin{figure}
	\includegraphics[width=1\linewidth]{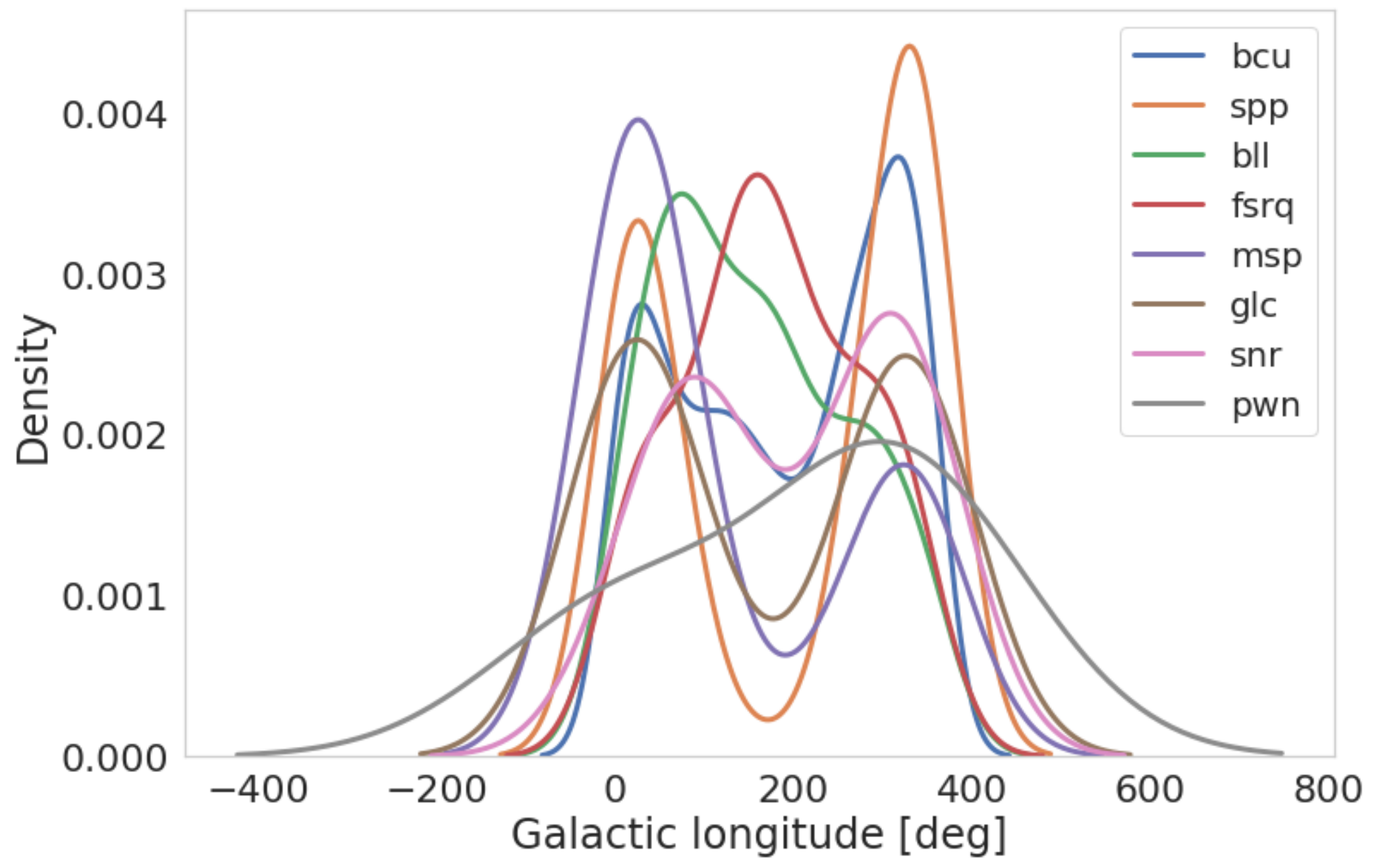}
	\vfill
	\includegraphics[width=1\linewidth]{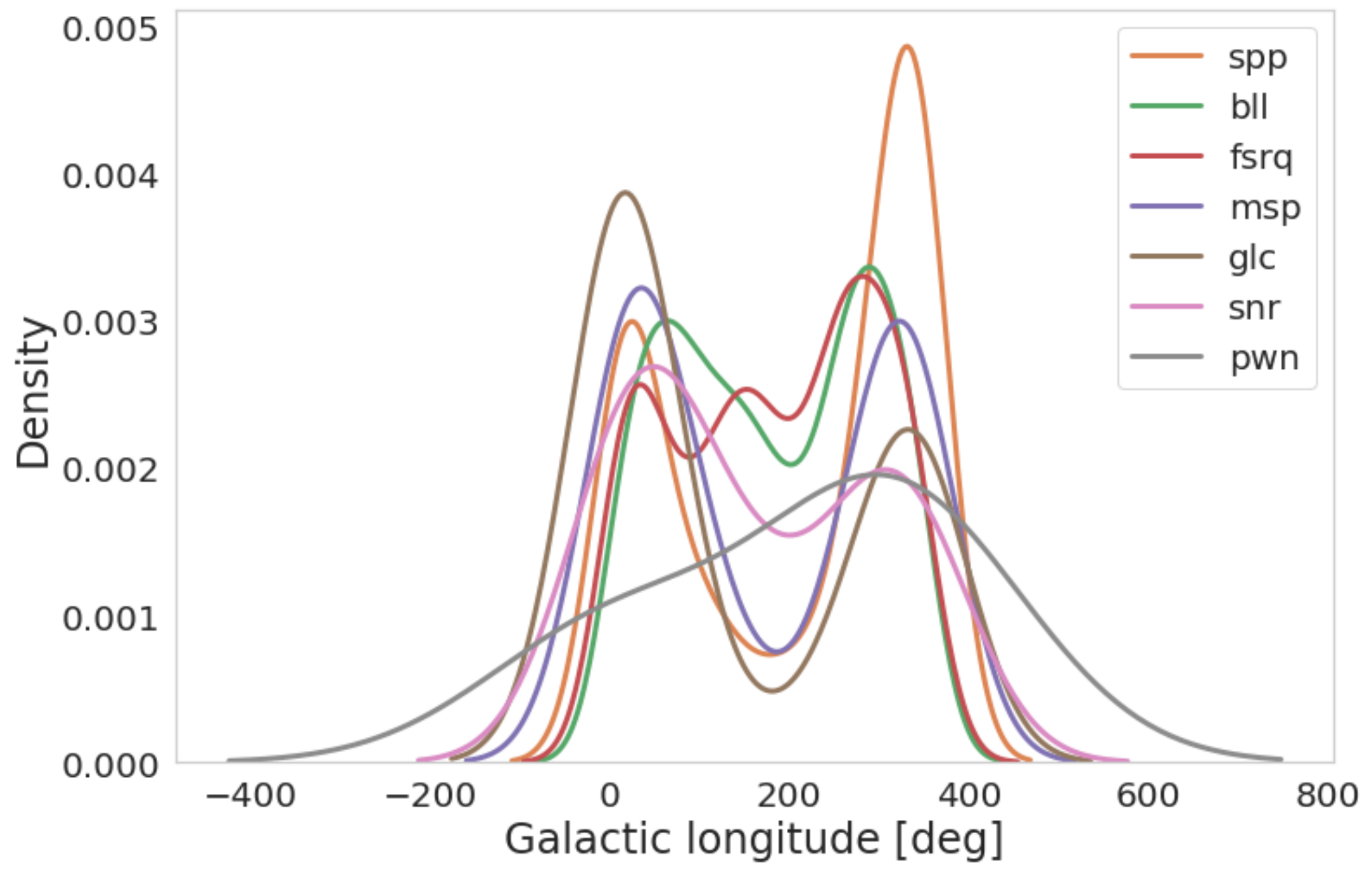}
	\vfill
	\includegraphics[width=1\linewidth]{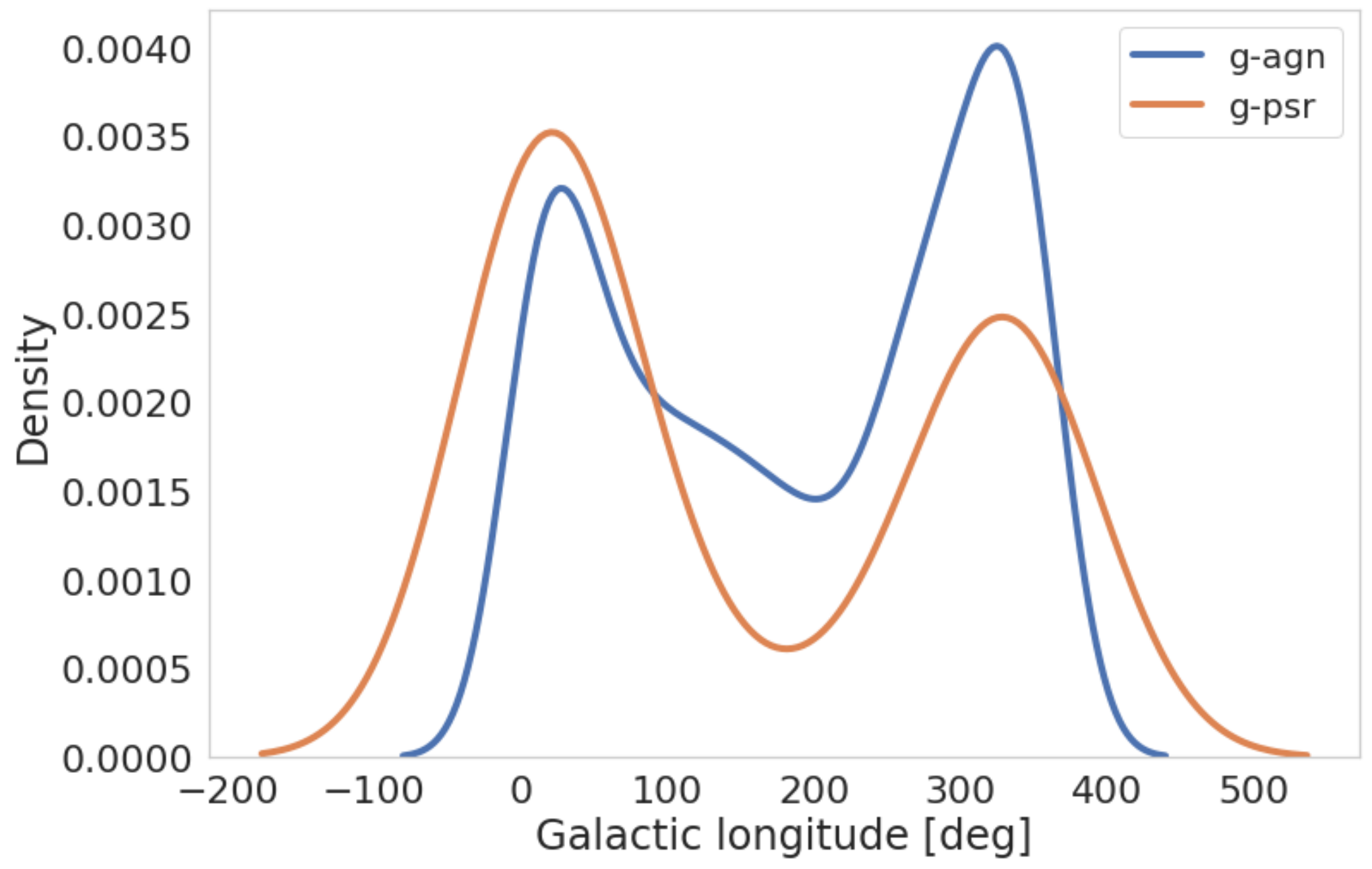}
   \caption{Galactic longitude probability distribution of the predicted classes in each of the three models. \textbf{Top row :} 23--class model. \textbf{Middle row:} 22--class model. \textbf{Bottom row:} 2--class model. As the Figure depicts the mathematical probabilistic density function, it may overcome the physical range [0, 360].}
    \label{fig:pred_longitudes}
\end{figure}

Regarding the Galactic longitude/latitude distribution, we plot them in Figures \ref{fig:pred_latitudes} and \ref{fig:pred_longitudes} for each of the three models. Only classes with more than one predicted member appear. As the unIDs are clustered around low latitudes, the predictions will also be; nevertheless, we find interesting differences depending on the predicted class and model.

All three models present distributions peaked around low latitudes, where most of unIDs are present, due to source confusion induced by the intense Galactic diffuse emission. It is interesting to note that, in the binary model, AGN-like sources are more spread across higher latitudes, while PSR-like's probability distribution decreases quicker towards high latitudes. This is consistent with the latitude distribution of known sources, where psr-like sources cluster around low Galactic latitudes, while agn-like sources are distributed more homogeneously all across the sky \cite{Abdollahi_2020}.

In the 22-class model, without BCUs, we note a symmetry of the distributions with respect to the central latitude, with the two predicted PWNs very peaked at zero latitude. Nevertheless, in the full, 23-class model, we note BLL and FSRQ distributions are shifted towards positive latitudes.

Regarding the longitude distribution, most classes present a double-peaked distribution with a local minimum around the Galactic anti-center (180 deg), as expected. There are some exceptions to this, such as FSRQs and BLLs, which do not decay towards the anti-center as they are extragalactic sources and do not follow a Galactic distribution. The eccentric distribution of PWNs is due to be composed only of three sources in both 23- and 22-class models.

The full prediction tables for each unID and model including the source name, Galactic coordinates, and predicted class are available as interactive arrays in the provided notebook.

\section{Clustering of associated and unID populations of gamma-ray sources}
\label{sec:clustering}
The classification of sources requires a label, i.e., a target variable. Another possible approach is to perform an unsupervised learning of the LAT gamma-ray sources, known as clustering. In this case, no source class information is required, as the algorithm will assign each source to different clusters, maximizing the intra-cluster similarity and differences between clusters.

This blind search will allow us to compare both associated and unID samples to search for possible statistical differences between them. If an extra cluster was found in the case of the unIDs, it would point towards the potential existence of a new class of gamma-ray emitter, difficult to associate to a known source in other wavelength and therefore only present in the unID sample. Therefore, no classification is attempted in this Section, but rather a statistical study of both associated and unID populations in order to compare their similarities.

To do so, we will use the k--means method \cite{1056489}. This is preferred over hierarchical, agglomerative clustering \cite{2010arXiv1012.3697A} due to the number of features considered, as well as the high number of individual sources. The main input is the number of clusters, which is not decided by the algorithm, but defined by the user.

We use the \texttt{scikit-learn} \cite{2012arXiv1201.0490P} implementation of k--means with \texttt{k-means++} initialization\footnote{See \href{https://scikit-learn.org/stable/modules/generated/sklearn.cluster.KMeans.html}{scikit-learn} webpage for more details on this method.}. The same features as in Section \ref{sec:classification} are removed, plus the \texttt{'name'}, \texttt{'source\_type'} and categorical variables, resulting in a 32--dimensional clustering. We divide the LAT sources in associated and unIDs, as our main goal in this section is to compare the number of predicted clusters, which could be interpreted as groups of sources with similar characteristics (not necessarily belonging to the same classes).

\begin{figure}
	\includegraphics[width=\linewidth]{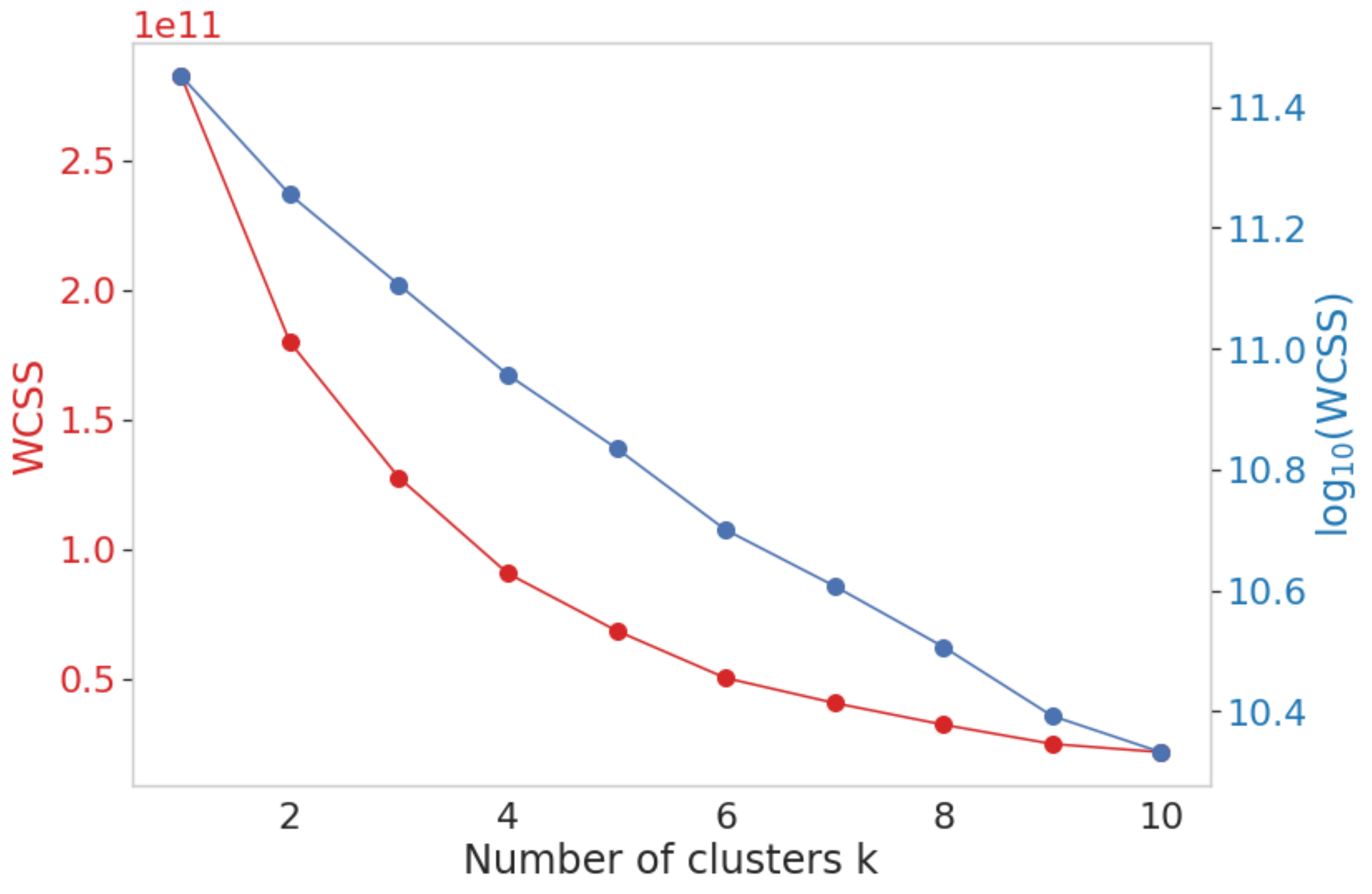}
	\vfill
	\includegraphics[width=\linewidth]{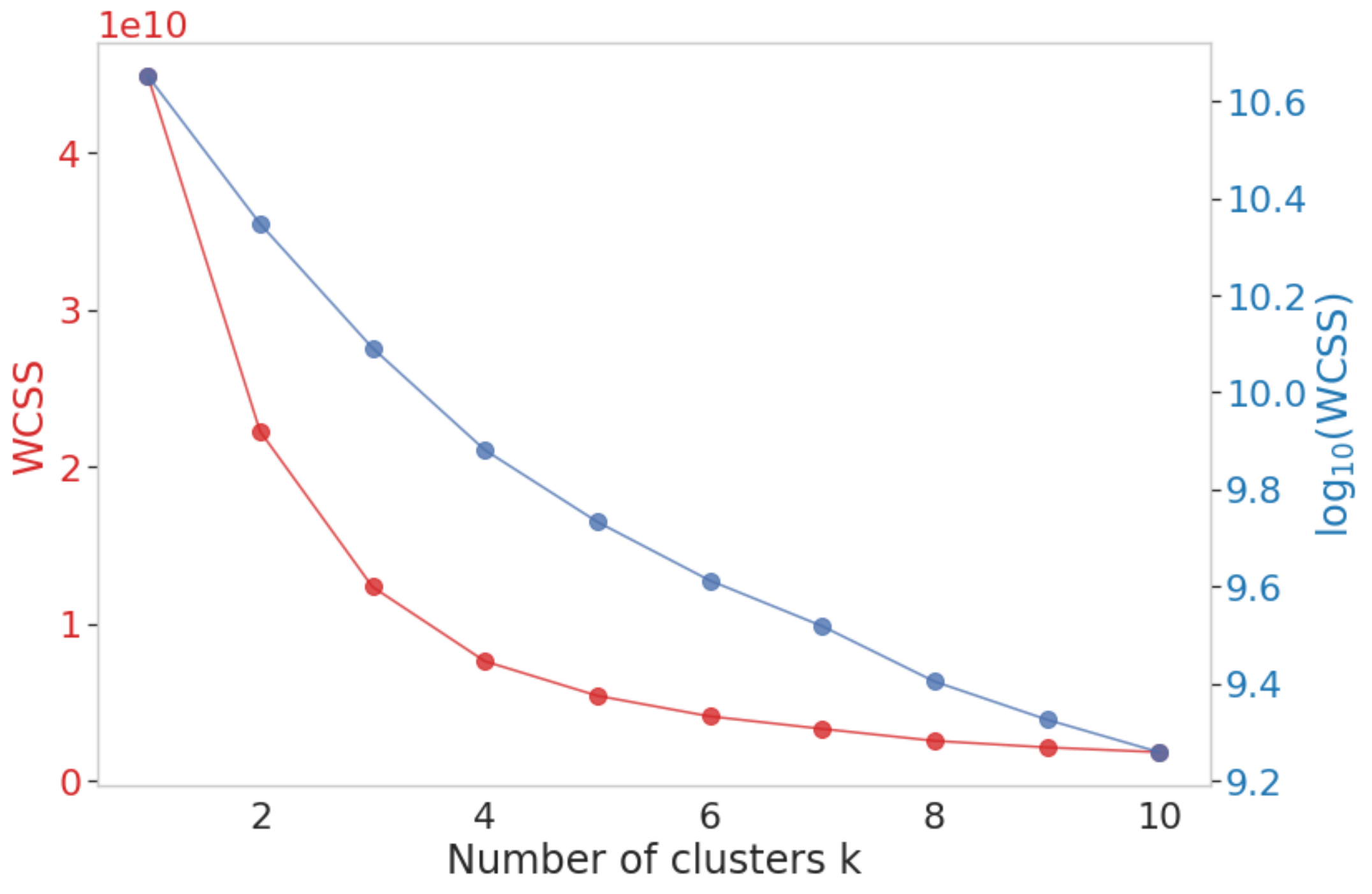}
   \caption{Within-Cluster Sum of Squares (WCSS) of k--means clustering on LAT sources. Top (bottom) panel show the result for the associated (unID) population. We show the WCSS curve in linear and log-scale to better see the ``elbow''.}
    \label{fig:elbow_kmeans}
\end{figure}

To decide the optimal number of clusters $k$, we implement the Within-Cluster Sum of Squares (WCSS) and the so-called ``elbow method'' \cite{8549751}. It consists on computing the sum of squares of the distances of each source in all clusters to their respective centroids. This will output a monotonically decreasing function, which will be null in the limit of $k=d$, where $d$ is the number of sources. Nevertheless, by evaluating the improvement on WCSS as we increase $k$, the curve will show an ``elbow'' on the optimal value $k_{opt}$, and for $k>k_{opt}$ the decrease will be less pronounced. The result of the WCSS curve for LAT sources clustering in both populations is shown in Figure \ref{fig:elbow_kmeans}.

In the case of the associated sample, it is hard to decide whether 4 or 6 is the optimal number of clusters with the red curve (in linear scale). Yet, by drawing it in log-scale, the value of $k_{opt}=6$ is preferred over 4. On the other hand, the unID WCSS curve has a clear ``elbow'' for $k=4$. Analytically, the computation of $k_{opt}$ is obtained computing the second derivative of a spline interpolation of the WCSS: the lowest $k$ in which the second derivative drops near zero will be $k_{opt}$. The values obtained this way match the aforementioned ones.

The possible mismatch between the predicted clusters on associated sources and unIDs suggests the former contains a few unique sources not present in the latter. By taking a look at the distribution of sources in each cluster, for both populations, we obtain 3902--303--142--13--2--1 ($\sim$89\%--7\%--3\%--0.3\%--0.05\%--0.02\%) for the associated sample and 1993--272--27--4 ($\sim$87\%--12\%--1\%--0.2\%) for the unIDs.

If we used $k_{opt}=4$ in the associated sample, the proportion would be 4324--30--2--1, with the two intermediate clusters of 303 and 142 sources being incorporated to the 3902- and 13-source clusters. The two last, exotic clusters are still present and therefore are not a result of over-estimating the $k_{opt}$.

Indeed, the two-member cluster is composed by the Vela \cite{2010ApJ...713..154A} and Geminga \cite{2020A&A...643L..14M} pulsars, among the first detected gamma-ray sources due to their proximity to the Solar System. The single-member cluster contains 4FGL J1533.9--5712e, an extended source classified as SNR with H.E.S.S. counterpart \cite{2017ApJ...843...12A}, which the k-means algorithm renders as unique. These are exceptional sources which are very easily associated, and not expected in the unIDs pool. Without these three rare sources, we would have four clusters with similar proportions (as discussed in the previous paragraph) in both the associated and unID samples, suggesting their underlying populations are statistically the same, according to the resulting gamma-ray spectral characteristics.

Additionally, we show in Table \ref{tab:summary_clusters} the distribution of source classes across all 6 clusters for the known sample. Clusters \#0 and \#1 present similar proportion of sources, yet the first does not contain g-psr sources, but the latter contains more than an order of magnitude more sources. Clusters \#3 and \#4 present similar distribution of sources among them, more balanced than \#0 and \#1, and again the latter presents a factor $\sim$10 more sources than the former. Finally, clusters \#2 and \#5 are the two- and one-member aggregations already discussed.

\begin{table}
\centering
\begin{tabular}{ |c|c|c|c|c|  }
\hline
\rowcolor[gray]{.8} 
\multicolumn{5}{|c|}{Cluster source class distribution} \\
\hline
\hline
\multicolumn{1}{|c|}{Cluster} & Members & AGN-like & PSR-like & Other \\
\hline
0 & 303 & 85\% & 0\% & 15\%\\
\rowcolor[gray]{.9} 
1 & 3902 & 87\% & 6\% & 7\%\\
2 & 2 & 0\% & 100\% & 0\%\\
\rowcolor[gray]{.9} 
3 & 13 & 46\% & 38\% & 16\%\\
4 & 142 & 57\% & 26\% & 17\%\\
\rowcolor[gray]{.9} 
5 & 1 & 0\% & 0\% & 100\%\\
\hline
\end{tabular}
\caption{Distribution of the source class distribution in the K-means clusters for the known sample.}
\label{tab:summary_clusters}
\end{table}

\begin{figure}
	\includegraphics[width=\linewidth]{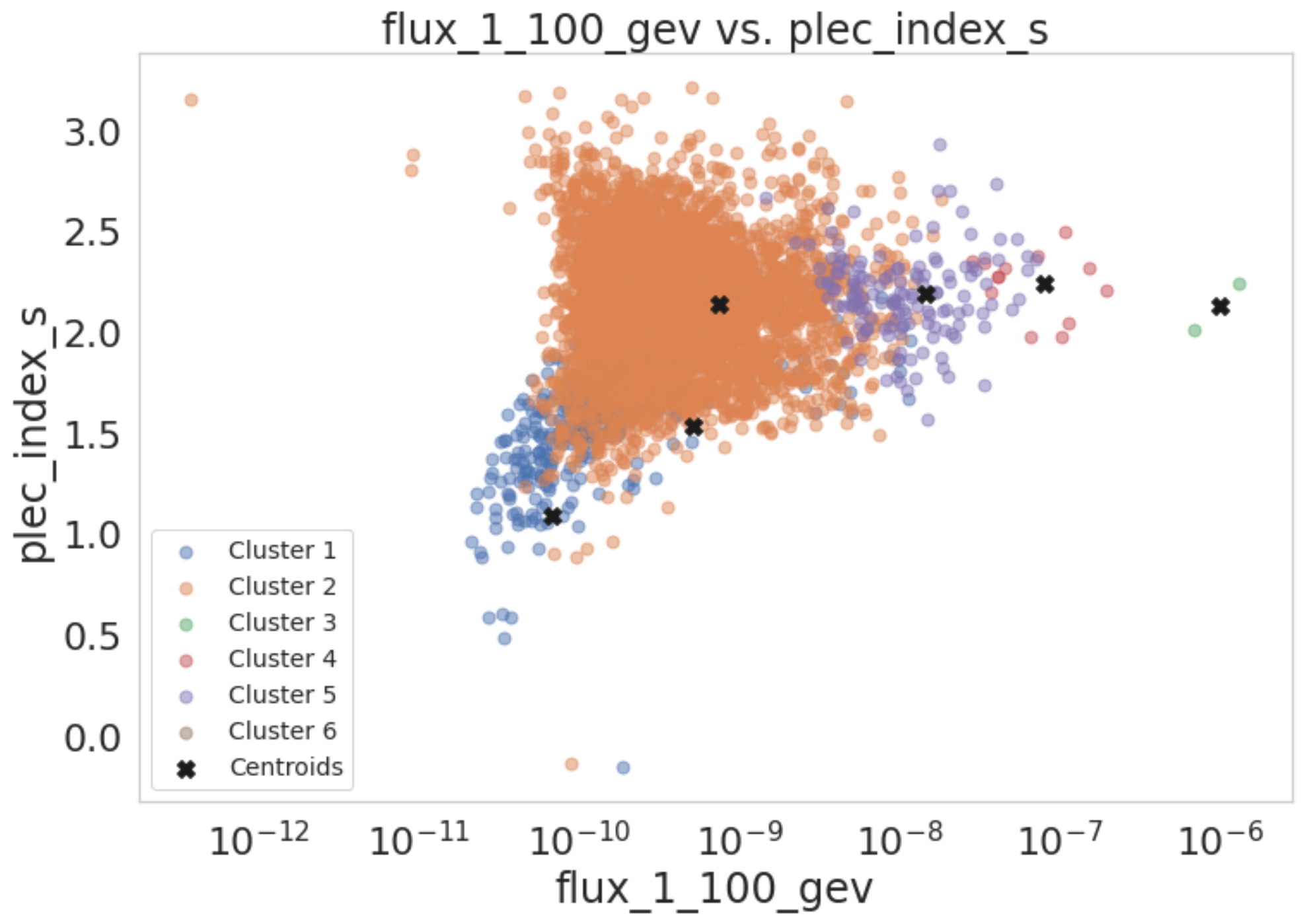}
   \caption{Example of clustering, for the associated sample $\left(k_{opt}=6\right)$. Colors indicate different clusters, while black crosses mark their centroids. The two green points at the right are the Vela and Geminga pulsars, which constitute the 2-member cluster. See the notebook for all possible combinations.}
    \label{fig:example_cluster_plot}
\end{figure}

We refer the interested reader to the full notebook to find a 2D plotting function which takes all possible combinations of variables and plots them against each other, with individual sources colored by cluster and the centroids marked in black. As an example, we show in Figure \ref{fig:example_cluster_plot} the scatter relation between two variables for the associated sample.

\section{Conclusions}
\label{sec:conclusion}

In this paper, we have performed a classification of \textit{Fermi}-LAT gamma-ray unidentified sources (unIDs) with machine learning techniques. We used the latest and most complete gamma-ray catalog released, the 4FGL--DR3, which contains more than 6000 sources, of which around 1/3 remain as unIDs.

To do so, we trained \texttt{CatBoost}, a state-of-the-art classification algorithm based on gradient boosted decision trees, with the associated sample of the catalog. By doing a 75/25\% train/test split and performing a 5-fold cross-validation on the test set, we obtain an average accuracy of 67\% of a 23-type classification. The largest source of confusion comes from the BCU class, which is itself a misclassification as they are blazars of unknown type.

We develop two alternative models to tackle this confusion: by removing BCU from the train pool, the accuracy is improved up to 81\% on average, while grouping 93\% of the catalog in PSR-like and AGN-like sources yields a binary classification problem with 99\% average accuracy. In all three models we perform a 5-fold cross validation to ensure proper variance in the selection of the test sample, finding a very narrow dispersion in the accuracies.

With these three models, we predict the classes of the unID sample. In the 23-class model, predictions are dominated by BCU, followed by SPP and BLL. A similar trend is found in the 22-class model, where the ranking is SPP, BLL and FSRQ. Finally, when predicting binary classes for AGN-/PSR-like sources, we find an agreement between the proportions on the both the associated set and the unIDs predictions.

\cite{2016ApJ...820....8S} performed a search for pulsars using RF and LR. In that work, authors used the 3FGL catalog, which comprised 4 years of Pass 7 LAT data and an outdated diffuse emission model. It is not possible to establish a fair comparison with our work, given that both algorithms and datasets are different  -- in our case, with 12 years of Pass 8 LAT data and a new diffuse model, the spectral characteristics of unIDs (which are by nature faint sources) will be altered. Nevetheless, by using our 22-class model, we cross-check our results for the msp class with Table 12 of \cite{2016ApJ...820....8S}, finding 36\% of our sample in the latter.

The non-coincident 64\% may be due to i) 3FGL dim sources which now fall below the detection threshold with more years of data, Pass 8 event reconstruction and the state-of-the-art diffuse model; ii) a different classification due to altered spectral properties with the 12-year data; or iii) different prediction outcomes due to the different algorithms used. The interested reader may use the provided Python notebook to search for any possible source of interest and compare the predicted class to the one of \cite{2016ApJ...820....8S} or any other work.

We can also study the spatial distribution of each class in Galactic longitude and latitude, finding a common trend: classes are low-latitude (as unIDs are) and decrease towards the Galactic anti-center. We find exceptions to this rule, most notably FSRQ and BLL, which are extragalactic sources. 

This classification can be useful for several science cases and statistical population studies, as well as planning multiwavelength campaigns.

Moreover, we perform an unsupervised learning on both associated and unID sets, clustering the sources according to similar characteristics across 32 variables with the k-means algorithm. Six clusters are found the associated sample, and four in the unID one. While this mismatch may be surprising, we show the two extra clusters of the associated sample are due to three exceptional sources, the Geminga and Vela pulsars, and an extended source detected with H.E.S.S. If we ignored these objects, both samples would have four clusters with a similar proportion of sources in each one.

This may be discouraging, as no exotic source --such as WIMP dark matter (DM) annihilation or decay \cite{2014arXiv1411.1925C, 2015PNAS..11212264F}-- would be expected to appear within the unIDs. Yet, the clustering results only tell us that there seems to be no extremely unique unID, but says nothing about their underlying physics. For example, low-mass DM annihilation in certain hadronic channels such as $b\bar{b}$ or $c\bar{c}$ can mimic PSR spectra \cite{Mirabal:2013rba, Mirabal:2016huj, 2019JCAP...11..045C}. Therefore, even if not exhibiting any exotic characteristic according to the 32 considered variables, it would constitute a new class of gamma-ray source.

\section*{Acknowledgements}

The author would like to thank the Fermi-LAT Collaboration for the public availability of data. This work made use of NASA’s Astrophysics Data System for bibliographic information.

\section*{Data Availability}
\label{notebook}

The datasets in csv and Excel format, and the full code to obtain the results presented in this paper can be obtained from \href{https://drive.google.com/drive/folders/1ZWGbMYQMQr60kOn9OhOR8SyvifVRvLmQ?usp=sharing}{this link}, as well as online supplementary material from MNRAS. The code is written in Python as an interactive, pre-computed notebook which can be opened and executed with \href{https://colab.research.google.com/}{Google Colab} or \href{https://jupyter.org/}{Jupyter}.



\bibliographystyle{mnras}
\bibliography{References_catboost} 





\bsp	
\label{lastpage}
\end{document}